*SCHOOLS ON DIFFERENT CORNERS: An Investigation into the Effects of Ethnicity and Socioeconomic Status on Physics Offerings in Northern California Public High Schools.*

David Marasco[1] and Bree Barnett Dreyfuss[2]
[1]Foothill College, Los Altos Hills, CA 94062
[2]Amador Valley High School, Pleasanton, CA 94566

## INTRODUCTION

In the spring of 2018 the Northern California/Nevada section of the American Association of Physics Teachers was alerted to a local high school's plans to eliminate physics for the following school year. As part of the campaign to support the school's efforts to sustain physics in the following year, the physics offerings from the surrounding schools in that district were compiled. It appeared that the demographics of the student population in the district played a role in the number of different physics courses offered within that district, particularly the percentage of Hispanic students (%Hispanic) and percentage of socioeconomically disadvantaged (%SED) students at each school. Concerned that this trend was more widespread, physics course offerings were reviewed for Northern California public high schools to determine if there were correlations between the amount of different physics class offerings and these populations. It was found that %Hispanic and %SED are strongly correlated in California public schools, and along with number of students, could be used as statistically significant predictors of a school's physics offerings.

## LITERATURE REVIEW

Physics enrollment in high schools is influenced by the access to physics courses, of any level, at each school. Physics availability in the U.S. has been studied by the American Institute of Physics (AIP) since the 1986-1987 school year, and they have found that each year over 90% of seniors attend a school that offers at least one physics course regularly.[1] AIP further found that California had above-average availability when compared to other states due to its larger schools and larger amounts of high school graduates.[1] However, our data suggest that the access to physics courses is not equitable across all ethnic groups or socioeconomic levels.

It is expected, according to the National Center for Education Statistics, that by 2023 nearly 30% of all school age (early childhood through grade 12) enrollments across the U.S. will be Hispanic.[2] In California specifically, the percentage of the state's population that identifies as Hispanic or Latino is much higher. Over 54% of public school students in the 2017-2018 school year designated their ethnicity as Hispanic or Latino.[3] Despite an increase in Hispanic student populations over the decades, Hispanic students have had consistently lower enrollment in high school physics classes throughout the United States.[4] According to the National Science Foundation High School Longitudinal Study of 2009, analyzed in a 2018 report, barely 40% of high school Hispanic students enrolled in a physics class.[5] A U.S. Department of Education study for the 2013-2014 school year found that 67% of schools with low black and Latino enrollments (less than 25%) offered physics, while only 48% of schools with high black and Latino enrollments (more than 75%) did.[6]

In an AIP study, high school principals were asked to classify their students as either "worse off," "average" or "better off" than students in their surrounding area.[4, 7] This differs from many other studies that base the number of SED students on the percentage that qualify for a school's free and reduced lunch program.[7] The AIP study found that of all students that did not have access to physics across the U.S., over half of them were in worse off schools compared to only about 10% of the students in better off schools that did not have access to physics.[4, 7] After isolating enrollment in higher-level physics courses, AIP found that 40% of students at better off schools, one third of those at average schools, and about 25% of students at worse off schools enrolled in honors or Advanced Placement ® (AP) or some other second-year physics course.[4, 8] They found that across the levels of physics available, "As the socioeconomic status increases, the proportion of students enrolled in higher-level physics classes increases."[4] This agrees with the NSF longitudinal study, analyzed in 2018, which found that 32% of the lowest quintile of students according to their socioeconomic status and over 56% of the highest quintile took a physics course of any level.[5]

According to the AIP study, Hispanics and blacks are least likely to attend a better off school, and twice as likely to attend a worse off school.[4, 9] With more Hispanic students attending schools designated as worse off compared to other students in their immediate areas, fewer have access to physics courses in high school.[4, 10]

**DATA COLLECTION**
The California Department of Education has demographic data including ethnicity and socioeconomic status as part of their School Accountability Report Card (SARC).[11] The data reported reflect the 2016-2017 academic year. The sample was restricted to Northern California, arbitrarily defined as the counties above the dividing lines between Monterey, Kings, Tulare and Inyo counties to the north, and San Luis Obispo, Kern and San Bernardino counties to the south (Fig. 1). This separation would include major cities in Northern California including San Francisco, San Jose, Oakland, Sacramento and Fresno while excluding major Southern Californian cities such as Los Angeles and San Diego. As this is a study concerning access, the sample was further limited to public, non-charter high schools, as these represent institutions where students would not have to either pay for their education, or possibly win a lottery to enable access. Only schools with a student population above 200 were examined as class offerings are further restricted in smaller schools. Once these schools were selected, the possible physics offerings for each were determined by the University of California (UC) Course List search page,[12] to which offerings at the given high schools are reported to the University of California system by the schools themselves. For each school, the different physics offerings were recorded as conceptual physics, physics, honors physics, AP-1, AP-2, AP-C (mechanics), or AP-C (E&M). The UC Course List allowed for a data set that was both complete and uniform in labeling convention for that year. The physics offerings were listed for the 2018-2019 academic year, but while a physics course may be listed, and thus may appear in a school's course catalog, it may not be offered each year. Such information is outside the scope of this paper. There were 410 high schools that met the above-mentioned criteria for which data were collected. The schools were then coded to reflect four different cases of access: no physics offering at all, physics with no AP options, only one AP course offered, or multiple AP courses offered (meaning multiple different AP courses, eg. AP1 and AP2, as opposed to multiple sections of one level of AP).

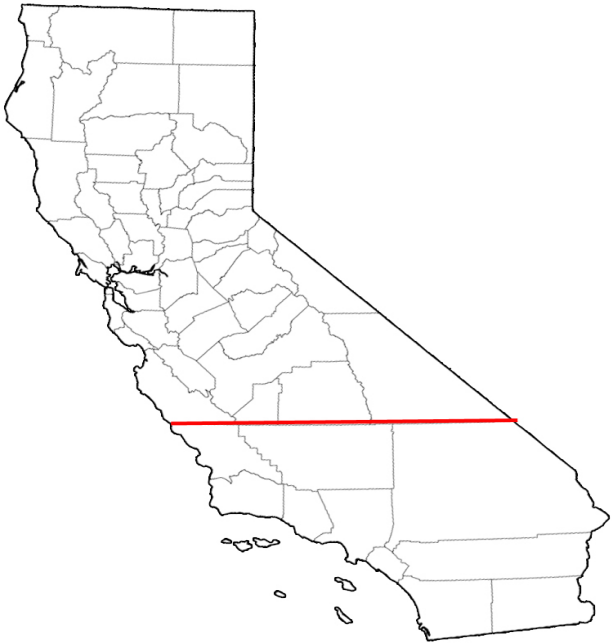

*Counties included:*
Alameda, Alpine, Amador, Butte, Calaveras, Colusa, Contra Costa, Del Norte, El Dorado, Fresno, Glenn, Humboldt, Inyo, Kings, Lake, Lassen, Madera, Marin, Mariposa, Mendocino, Merced, Modoc, Mono, Monterey, Napa, Nevada, Placer, Plumas, Sacramento, San Benito, San Francisco, San Joaquin, San Mateo, Santa Clara, Santa Cruz, Shasta, Sierra, Siskiyou, Solano, Sonoma, Stanislaus, Sutter, Tehama, Trinity, Tulare, Tuolumne, Yolo, and Yuba.

*Counties excluded:*
Imperial, Kern, Los Angeles, Orange, Riverside, San Bernardino, San Diego, San Luis Obispo, Santa Barbara, and Ventura.

*Fig. 1. Map showing the geographical boundaries of the sample.*

DATA ANALYSIS
The schools were plotted as shown in Figure 2, with %Hispanic of the school population on the x-axis and %SED on the y-axis. There is a clear correlation between the two variables, with an r2 value of 0.60. As the %Hispanic and the %SED of the overall school population increases, the degree of physics access drops, with fewer schools offering multiple different AP courses, or offering any AP course at all. Of all the public, non-charter high schools in Northern California included in the study, none that were more than 70% Hispanic offer more than one type of AP course. There are 68 such high schools in the region. Of the 50 least-Hispanic schools, 17 offered multiple levels of AP. Of the 50 most-SED schools, only 3 offer multiple levels of AP, contrasted with 21 of the 50 least-SED schools in the study.

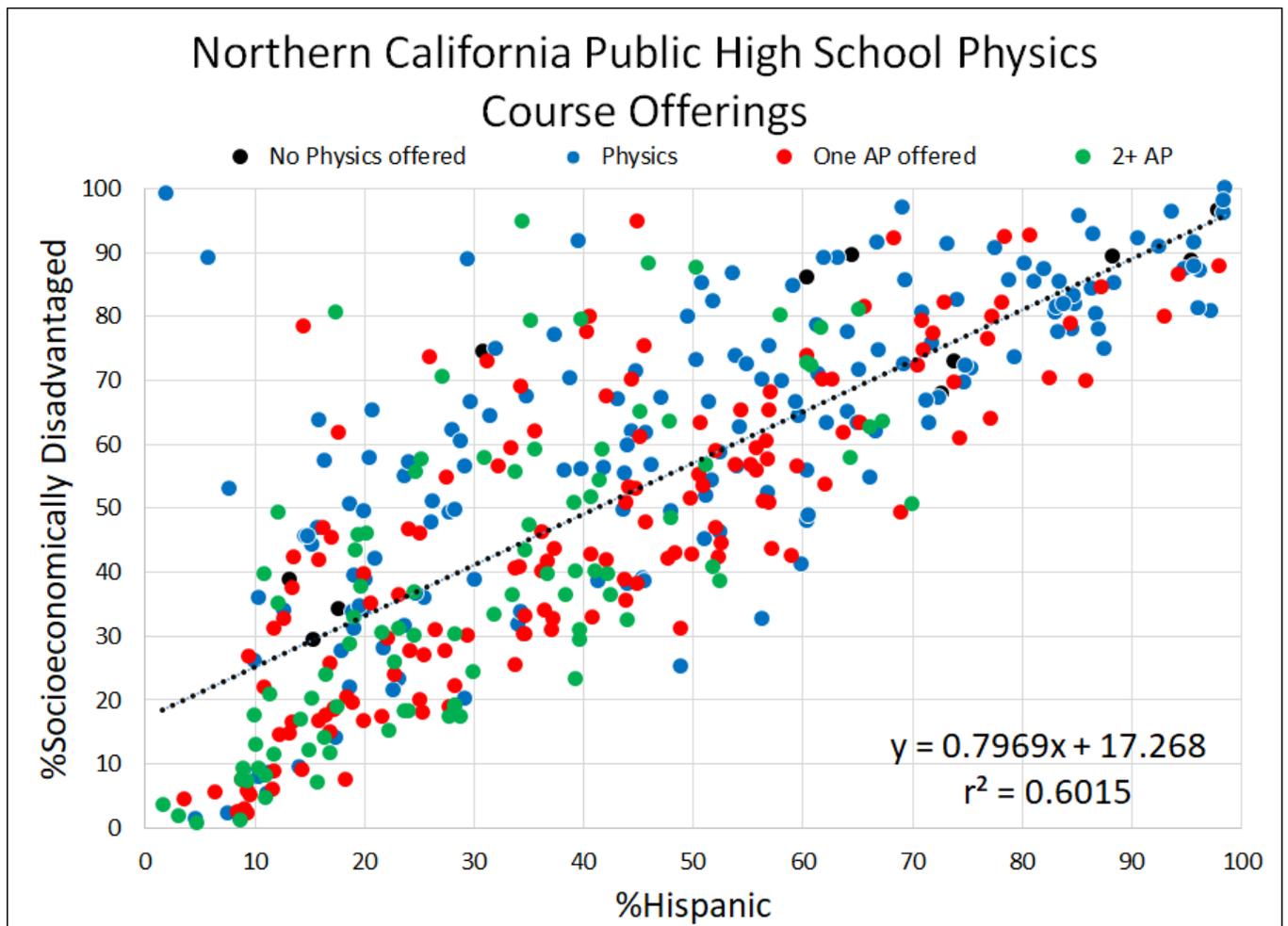

Fig. 2: The %Hispanic vs. %SED for Northern California public non-charter high schools greater than 200 students. Data are color coded for physics course offerings.

Of the 183 schools in the lower left corner (<50% Hispanic and <50% SED), three (1.6%) offered no physics whatsoever, 45 (24.6%) offered physics with no AP courses, 75 (41.0%) offered a single AP physics choice and 60 (32.8%) offered more than one type of AP class. In the upper right quadrant, there were 141 schools that were >50% Hispanic and >50% SED. In this grouping, seven (5%) offered no physics whatsoever, 80 (56.7%) offered physics with no AP courses, 43 (30.5%) offered a single AP physics choice and 11 (7.8%) offered more than one type of AP class. Note that 79.0% of the schools examined were in one of the two listed categories. In total, 73.8% of the schools in the lower left (<50% Hispanic and <50% SED) offered at least one AP course for their students, whereas only 39% of schools in the upper right (>50% Hispanic and >50% SED) did the same. These data are summarized in Table 1.

| Corner | No physics | Physics with no AP | Only one AP physics | Multiple AP physics |
|---|---|---|---|---|
| <50% Hispanic and <50% SED [n=183] | 1.6% [3] | 24.6% [45] | 41.0% [75] | 32.8% [60] |
| >50% Hispanic and >50% SED [n=141] | 5% [7] | 56.7% [80] | 30.5% [43] | 7.8% [11] |

*Table 1: Physics offerings for >50% Hispanic and >50% SED compared to <50% Hispanic and <50% SED*

Ordinal logistic regressions were run on the data, with no physics, just physics, AP physics and multiple-levels of AP physics as the result categories. As there were strong correlations between %Hispanic and %SED, regressions were performed on each of these measures independently in combination with the number of students enrolled in the school. It was found that the number of students served is a factor in what a school can offer. This result is consistent with earlier work.[13] This should not be surprising. In a model where students take physics in their senior year a school with 200 students will have a cohort of roughly 50 seniors, and it is unlikely that said school would have a wide selection of physics offerings. However, a school with 2000 students would have 500 seniors, and could support multiple levels of AP offerings.

Two logistic regressions were performed. In the first, number of students, %Hispanic, and an interaction term were evaluated. The interaction term was found to be not statistically significant, and was removed from the model. In the revised model, both number of students and %Hispanic were significant with p<0.001. The coefficient for number of students was 0.0015 with an odds ratio of 0.86 per 100 students. The coefficient for %Hispanic was 0.0273, with and odds ratio of 1.03. This analysis was repeated with %SED in place of %Hispanic, and once again the interaction term was removed as it was not significant. In this model student population and %SED were significant with p<0.001. The coefficients were 0.0013 for number of students and 0.0275 for %SED, with a corresponding odds ratio of 0.87 per 100 students and 1.03 for %SED. In both cases, %ELL (English Language Learners) was tested along with number of students and the appropriate demographic variable, and %ELL was not found to be a significant predictor.

**NEXT STEPS**
There are several paths for future work in this topic. Studies similar to this one should be performed looking at the availability of STEM AP classes along demographic lines. Are the access issues in the other fields consistent with the current findings in physics? Does the existence of one AP STEM class make others more likely at that school, and if yes, how so? Are there "AP STEM deserts" that reflect certain demographics? In 2017 Hispanic students took about 39% of the AP exams in the state of California, despite representing 54% of the state's public school population.[14] In addition, examination of access through a rural/suburban/urban lens is of interest.

*Acknowledgements - The authors would like to thank Elaine Kuo, Doreen Finkelstein and Ben Stefonik for fruitful discussions, Brian Robinson for editorial feedback, and Angela Kelly for early encouragement. This work is dedicated to the students at James Lick High School.*